\documentstyle[sprocl]{article}

\input psfig.sty

\def\be{\begin{equation}}
\def\ee{\end{equation}}
\def\bea{\begin{eqnarray}}
\def\eea{\end{eqnarray}}

\def\c12ag{$^{12}C(\alpha,\gamma)^{16}O$}
\def\xn13pg{$^{13}N(p,\gamma)^{14}O$}
\def\be7pg{$^7Be(p,\gamma)^8B$}
\def\o14{$^{14}O$}
\def\xbe7{$^7Be$}
\def\b8{$^8B$}
\def\n16{$^{16}N$}

\begin{document}

\thispagestyle{empty}

\hspace{3in} {UConn-40870-0022}

\title{How Well Do We Know the Beta-Decay of
  $^{16}N$ and \\ Oxygen Formation in Helium Burning
\footnote{Work Supported by USDOE Grant No. DE-FG02-94ER40870.}}

\author{Moshe Gai
\footnote{Invited Talk, Physics  
  With Radioactive Nuclear Beams, Puri, India, Jan. 12-17, 1998}}

\address{Dept. of Physics, U46, University of Connecticut,
   2152 Hillside Rd., \\ Storrs, CT 06269-3046, USA; 
   gai@uconnvm.uconn.edu, http://www.phys.uconn.edu}


\maketitle

\abstracts{We review the status of the \c12ag reaction rate, of 
importance for stellar processes in a progenitor star prior to 
a super-nova collapse.  Several attempts to constrain the p-wave 
S-factor of the \c12ag reaction at Helium burning temperatures 
(200 MK) using the beta-delayed alpha-particle emission of \n16 
have been made, and it is claimed that this S-factor is known, 
as quoted by the TRIUMF collaboration. In contrast reanalyses 
(by G.M. hale) of all thus far available data (including the \n16 
data) does not rule out a small S-factor solution. Furthermore, we 
improved our previous Yale-UConn study of the beta-delayed 
alpha-particle emission of \n16 by improving our statistical 
sample (by more than a factor of 5), improving the energy resolution 
of the experiment (by 20\%), and in understanding our line shape, 
deduced from measured quantities.  Our newly measured spectrum of the 
beta-delayed alpha-particle emission of \n16 is not consistent with the 
TRIUMF('94) data, but is consistent with the Seattle('95) data, as well 
as the earlier (unaltered !) data of Mainz('71). The implication 
of this discrepancies for the extracted astrophysical p-wave 
s-factor is briefly discussed.}

\section{Helium Burning: The \c12ag Reaction and the Beta-Delayed
      Alpha-Particle Emission of \n16}

The \c12ag reaction rate is of importance for understanding 
helium burning \cite{Fo84} in massive (red giant) stars. The result of 
helium burning is to create an (onion) layer composed of
oxygen and carbon.  The C/O ratio determines the elemental 
composition of the core of the progenitor star prior to 
a supernova collapse and it 
turns \cite{We93} out that a carbon rich star 
tends to leave behind a remnant 
neutron star, while an oxygen rich star leaves a black hole.

We study the \c12ag reaction in its time 
reverse process using the beta-delayed
alpha-particle emission of \n16, allowing us to add useful data 
and constraints on the reaction rate, and the
extraction of the p-wave astrophysical S-factor.
However, it appears that {\bf early hopes} for deducing the
p-wave astrophysical S-factor ($S_{E1}$) using the \n16
data {\bf are not substantiated}.  And further confusion is generated
by inconsistent data on the beta-delayed alpha-particle
emission of \n16 in addition to inconsistent data on the
\c12ag reaction.

We emphasize that while data on the beta decay of \n16 may
add useful constraint and may allow for extracting the (virtual) reduced 
alpha-particle width of the bound $1^-$ state, the sign of the
mixing phase of the bound and quasi-bound $1^-$ states in the
\c12ag reaction has nothing to do with the beta-decay of \n16 and
can not be directly determined from the data on the beta-delayed 
alpha-particle emission of \n16. It turns out that this difficulty
does not allow for unambiguous extraction of the p-wave S-factor
even with the inclusion of the new data on \n16.  Furthermore, 
a reanalysis of all existing data (including the \n16 data)
by Gerry Hale \cite{Ha96} demonstrates that a small S-factor 
solution could not be ruled out.  In fact Hale's best fit for the 
TRIUMF \n16 data \cite{Az94} is for an E1 S-factor approximately 
20 keV-b.  Interestingly, Hale's best fit is consistent with a broader 
line shape.  As we discuss below, such a broader line shape is observed 
in all other data sets, and it is quite possible that the 
narrow line shape of the TRIUMF data is an artifact of their data
analysis.

\subsection{The TRIUMF Result}

A measurement of the beta-delayed alpha-particle emission of \n16
was performed at TRIUMF \cite{Bu93,Az94}.  The spectrum is observed with
high statistics (approximately one million events) 
and indeed the TRIUMF collaboration claims to have deduced
the p-wave astrophysical S-factor with high accuracy.  
Based on for example, their R-matrix
analysis they quote a large value of:
$S_{E1}=81 \pm 21\ keV-b$. The E1 S-factor was 
previously uncertain by approximately a factor of 10 
and we note the relatively
high accuracy and the implication 
that they determined the interference of the two $1^-$ states in
\c12ag to be constructive (i.e. large S-factor).

As we demonstrate in this paper there is enough reasons to doubt the
TRIUMF data, and furthermore we do not confirm the conclusion of the
TRIUMF group that the p-wave S-factor of the \c12ag reaction has been
measured.

\subsection{The New Yale-UConn Experiment}

A further measurement of the Beta-Delayed Alpha-Particle energy 
spectrum of \n16 at low energy was performed in continuation of 
the first generation Yale experiment 
\cite{Zh93,Zh93a}.  The final phase of this experiment was performed 
using the Yale ESTU tandem van de Graaff accelerator at the Wright 
Laboratory at Yale University during the summer of 1995 \cite{Fr96,Fr96a}.

The \n16 was produced using a 70 MeV $^{15}N$ beam and a 1250 
Torr, 7.5 cm long deuterium gas target with 25 $\mu m$ beryllium 
entrance and exit foils.  The \n16 emerged from the gas target with a 
broad recoil energy spectrum, with the lower 1 MeV portion 
stopping in a thin (190 $\mu g/cm^2$) aluminum catcher foil tilted 
at 7$^\circ$ with respect to the beam.  After the \n16 was 
captured, the catcher foil was rotated 180$^\circ$ 
into the counting area.  While the arm rotated and the detectors 
counted, a tantalum beam chopper was used to block the beam far 
upstream.  Each full production and counting cycle lasted 21 
seconds, approximately twice the lifetime of \n16.

\centerline{\psfig{figure=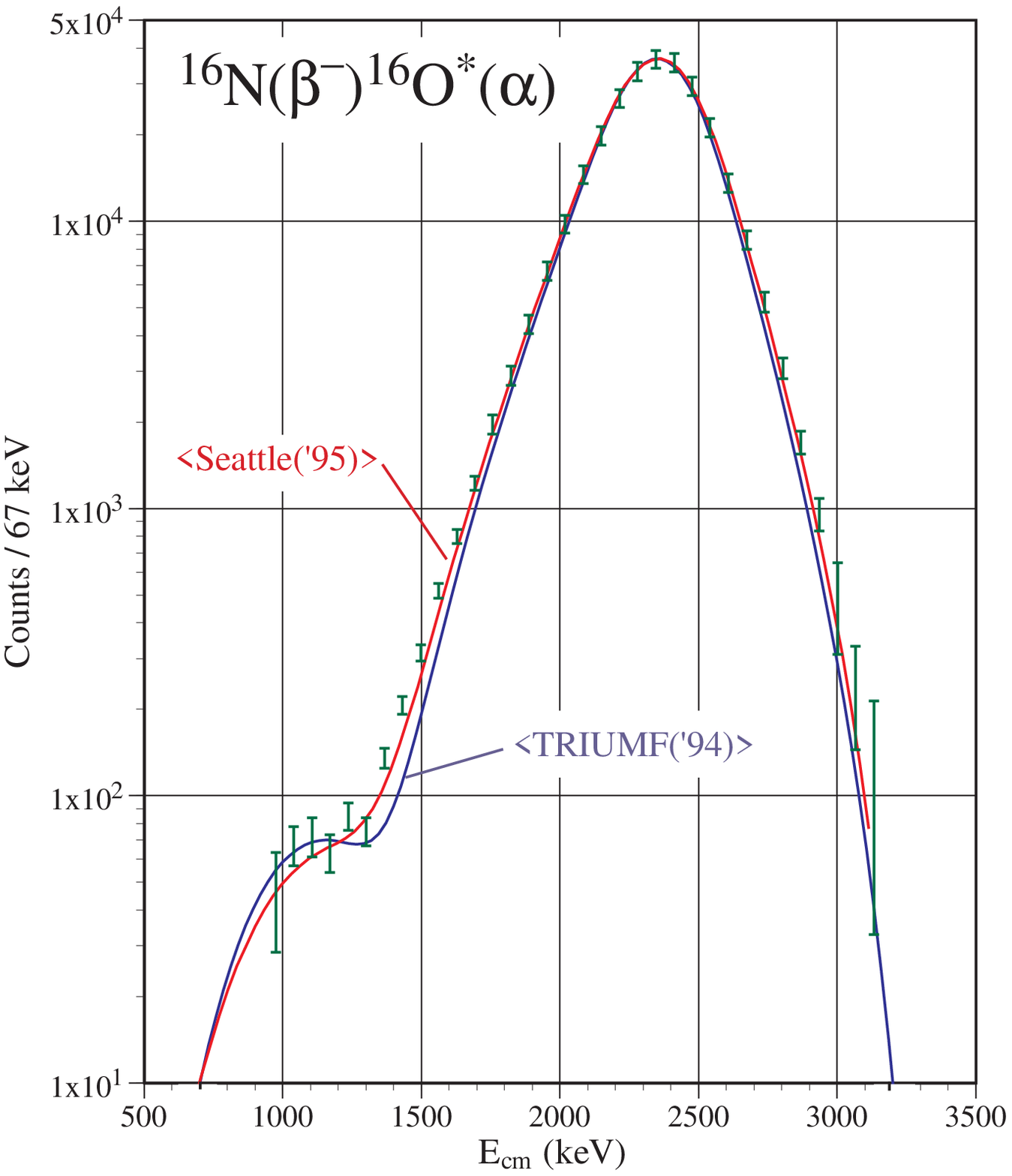,height=3.5in}}

\underline{Fig. 1:}  New Yale-UConn data, corrected for line shape,
compared with TRIUMF \cite{Az94} and Seattle \cite{Zh95} theory curves
(with reduced chi-squares).
The theory curves have been averaged over the 
experimental energy resolution.

The counting area contained, as in our previous
experiment \cite{Zh93,Zh93a}, 9 thin 
Silicon Surface Barrier (SSB) detectors used to measure the energy 
and timing information of the alpha-particles in coincidence with 
an array of 12 fast plastic scintillator detectors, which measured 
the timing of beta-particles.  This timing information was used to 
reduce (by more than a factor of 100 over the low energy range 
of interest) the background in our SSB array due to detection of 
beta-particles and due to partial charge collection in the SSB 
detector.

The line shapes of both the first and second Yale-UConn data sets
are the same \cite{Fr96,Fr96a}.
In order to consider the line shape of both 
Yale-UConn data sets, it is useful to consider a situation for a 
predicted spectrum which is constant in energy (or time).  Clearly 
the yield at a specific energy (time) is directly proportional to 
the energy (time) resolution at that energy (time).  In this case 
the energy (time) resolution is the integration interval.  
Hence our data need to be divided by the varying energy 
resolution for alpha-particles traversing our aluminum foil 
and the time resolution of our time of flight system. 
 The time resolution of our experiment is 
measured directly in the data on the beta-delayed alpha-particle 
of \n16 as well as the beta-delayed alpha-particle emission of 
$^8Li$ which was also measured in our experiment using the same 
setup and the $^7Li(d,p)^8Li$ reaction.  Hence the line shape in 
the current (and previous) 
experiment(s) is deduced from measured $\partial E/\partial x$ data 
and the measured time resolution of our experiment.

We improved our previous Yale-UConn experiment \cite{Zh93,Zh93a} 
by: (1) A 20\% improvement of our energy resolution (200 keV at 2.36 MeV), 
(2) More than a factor of five increase in statistics (292,000 
events), and (3) An understanding of our line shape deduced from 
measured quantities.  Our results are shown in Fig. 1.  The data 
shown in Fig. 1 were corrected for the energy dependence of the 
$\beta -\alpha$ coincidence efficiency and line shape, both
deduced from measured quantities. The uncertainty of the three
highest energy points include the uncertainty of the
$\beta -\alpha$ coincidence efficiency.

\subsection{Comparison of TRIUMF data to other data sets}

In Fig. 1 we also show our data compared to the Seattle \cite{Zh95} and 
TRIUMF \cite{Az94}\ theory curves 
averaged over the variable energy resolution of our
experiment.  Note that the theory curves are a good representation 
of their respective data, but they allow us to carry out the 
energy averaging also over the edges of the finite data.  With the 
Seattle theory superimposed on our data we calculate a
$\chi ^2$\ per data point of 1.4 and for TRIUMF theory 7.2.  We conclude 
that our data confirm the Seattle data \cite{Zh95}\ but 
do not confirm the 
TRIUMF data \cite{Az94}.  Most notable is the absence 
of a well defined minimum at approximately 1.4 MeV as suggested by 
the TRIUMF data.  The data in the vicinity of 1.4 is dominated by 
the f-wave contribution and hence essentially determines the f-wave 
contribution.  A larger f-wave contribution (at 1.4 MeV) would 
naturally lead to a smaller p-wave contribution at the interference 
maximum (at 1.1 MeV) and thus a smaller p-wave astrophysical S-factor.

\centerline{\psfig{figure=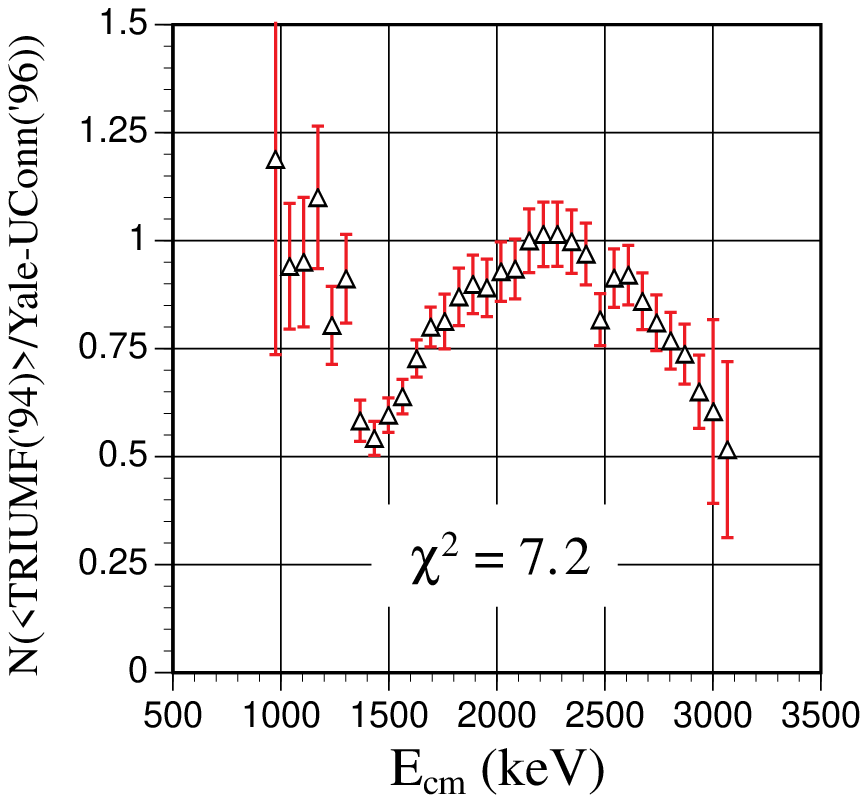,height=2.5in}}

\underline{Fig. 2:}  Ratio of the energy averaged TRIUMF theory 
curve  \cite{Az94} to the new Yale-UConn data.

\centerline{\psfig{figure=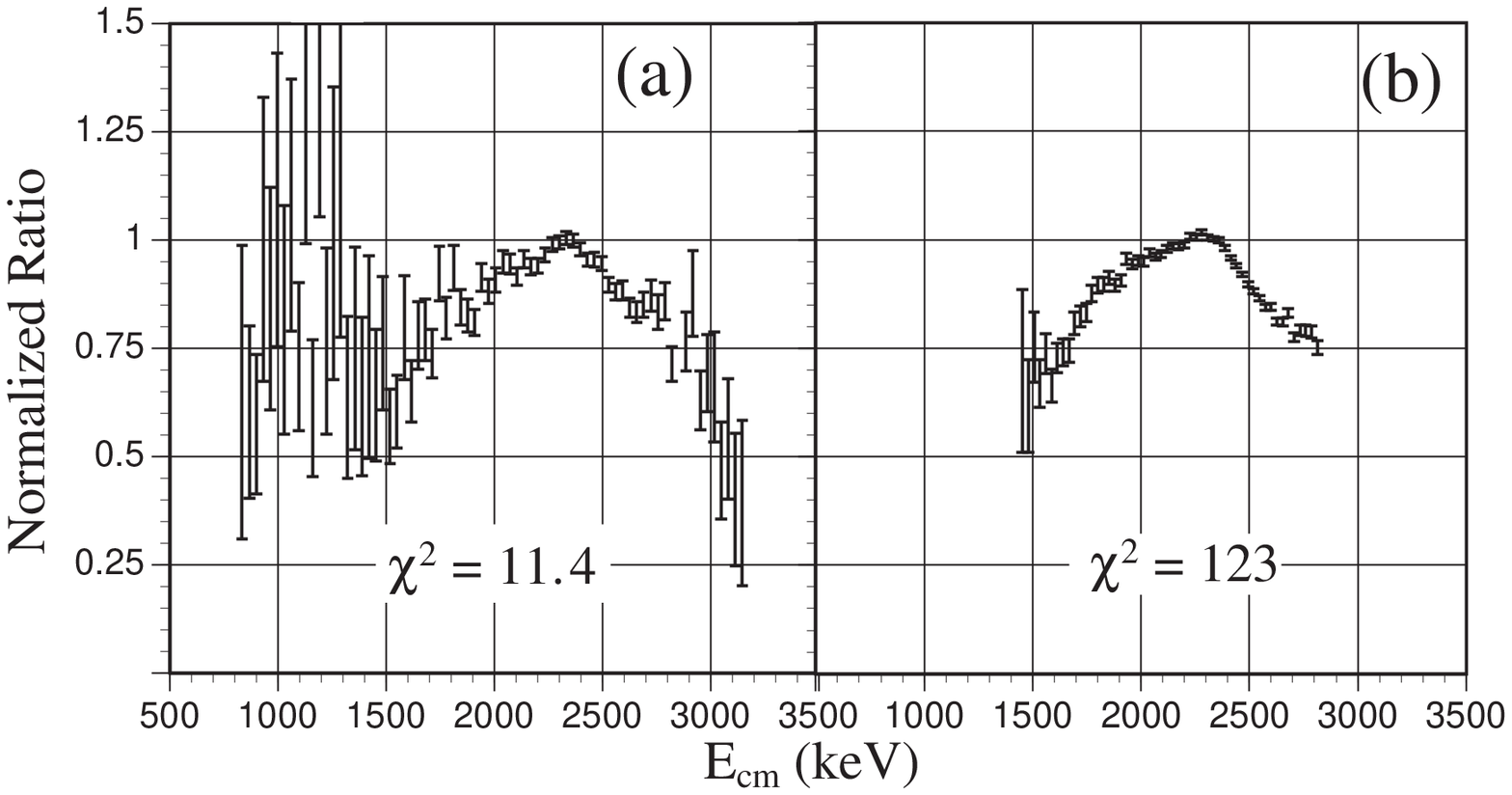,height=2.5in}}

\underline{Fig. 3:}  Ratios of (a) the TRIUMF data set \cite{Az94} to
the Seattle data set \cite{Zh95} and (b) the TRIUMF data set \cite{Az94} to
the (unaltered !) Mainz data set \cite{Ba96}.  
Linear interpolations were used
when necessary. Notice that the ratio plots are
very similar to each other and to the plot in Figure 2,
indicating that there are
three data sets: the Mainz('71), 
Seattle('95), and the current Yale-UConn('96) that agree with each 
other but disagree with the TRIUMF('94) data.

Following the conclusion that our data is consistent with the
Seattle data but not TRIUMF data, as shown in Figs. 1 and 2, 
we received from Fred Barker 
\cite{Ba96,Ha69,Ha70,Ne74} a copy of the original 
communication from Waffler to 
Barker dated 5 Feb. 1971, which includes approximately 32 million events 
and a measured beta-particle background spectrum. This data set was 
originally taken in a study of the parity violating alpha decay from the
8.8719 MeV $2 ^-$\ state in $^{16}O$.  We first note that we do not 
confirm \cite{Fra97} the allegation that there is a problem 
with the energy calibration of the Mainz data. We have in fact shown 
\cite{Fra97} that the alteration of the Mainz 
data by the TRIUMF calibration can 
not be justified. Using the original unaltered Mainz data
we observe that it agree with the Seattle data ($\chi ^2$\ per
data point of 2.5) and disagrees with the TRIUMF data ($\chi ^2$\ per 
data point of 123).  In Fig. 3 we show using a
linear scale, the ratio of the
TRIUMF(94) data to other data sets.  
Note that the disagreement with the 
TRIUMF data in all cases is equally bad on the high and 
low energy sides of the main peak at 2.35 MeV.  This together with the 
fact that all data sets agree on the low energy interference
maximum, negates arguments of low energy tails.  
We conclude that indeed all other
data sets that were measured with the \n16 produced via the 
$^{15}N(d,p)^{16}N$ reaction 
including Mainz(71), Seattle(95) and Yale-UConn(96)
agree with each other and exhibit the (same) disagreement with the
TRIUMF(94) data.

\subsection{Comparison of TRIUMF(93) data to TRIUMF(94)}

These disagreements suggest two possible conclusions.  One, that all data other
than the TRIUMF data are wrong and only the TRIUMF data exhibit the true
narrow line shape. Second, that the narrow line shape of the TRIUMF(94) data
is an artifact of the coincidence data analysis.

In order to further investigate these two possibilities we have examined
the TRIUMF(93) data \cite{Bu93} as compared to TRIUMF(94) data \cite{Az94}
-- as reanalyzed by the graduate student James Powell.
And in Fig. 4 we show the ratio of the TRIUMF(94) data to TRIUMF(93) data.
Clearly the TRIUMF(93) data exhibit yet even a narrower line shape than
TRIUMF(94). But the TRIUMF(93) data was already rejected by the TRIUMF
collaboration, as discussed by the 
collaboration \cite{Az94}, and clearly this demonstrates
that the narrow line shape of the TRIUMF(93) data is an artifact of the
analysis (i.e. energy miscalibration).

\centerline{\psfig{figure=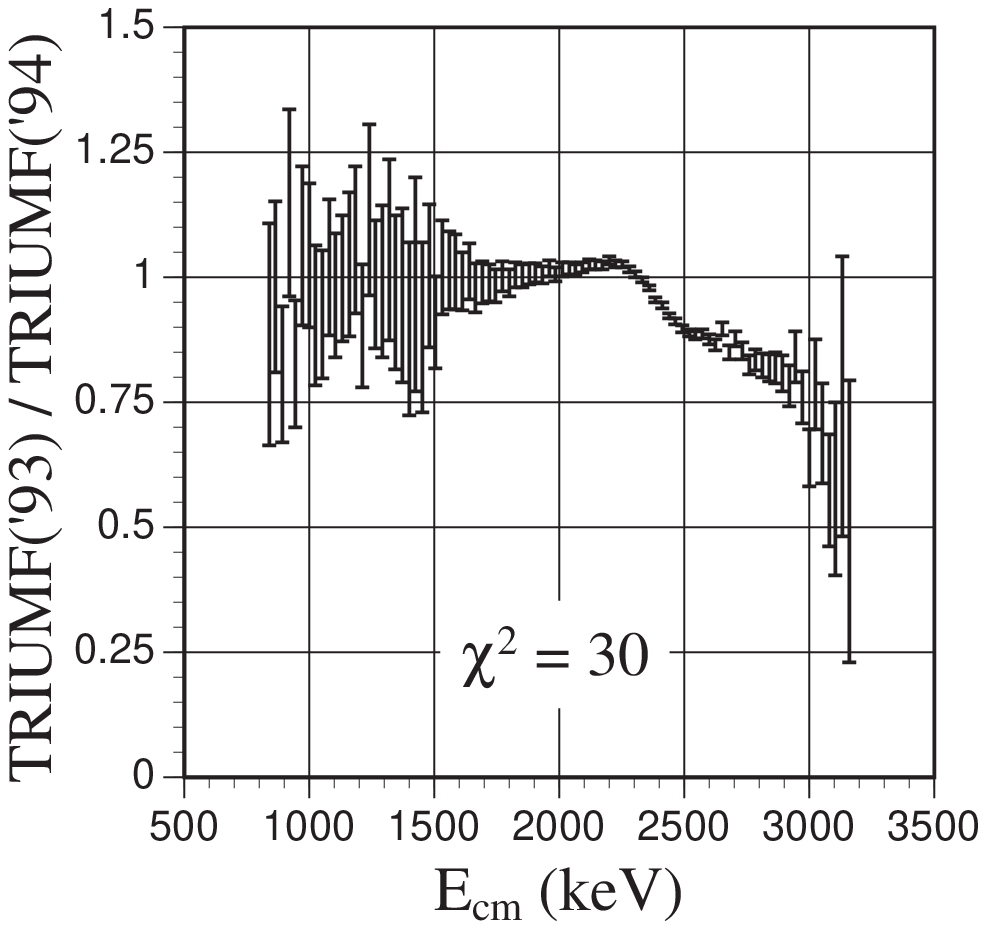,height=3.0in}}

\underline{Fig. 4:}  Ratio of the TRIUMF(94) \cite{Az94} to TRIUMF(93)
\cite{Bu93} data.  The narrower line shape of the TRIUMF(93) data is
understood to be an artifact of the analysis.

\subsection{Conclusions}

We have reviewed the status of both data and analyses pertaining to 
the p-wave astrophysical S-factor of the \c12ag reaction. We observe 
that more recent global R-matrix fit of 
the data on \n16, elastic scattering and \c12ag reaction data
does not allow us to rule out a small S-factor solution and does 
not confirm the strong statement of the TRIUMF collaboration that 
the S-factor is now known.  The sign of the 
interference of the two $1^-$ states in \c12ag data is not 
directly determined by data on the beta-decay of \n16,
and thus this problem remains unsolved and needs to be studied via
additional low energy data on the \c12ag 
reaction itself.

We have improved our original 
data on the beta-delayed alpha-particle
emission of \n16.  A comparison of all four high 
statistics data on \n16 reveals three data sets: the Mainz('71), 
Seattle('95), and the current Yale-UConn('96) that agree with each 
other but disagree with the TRIUMF('94) data (by up to 
a factor of 3). The current situation
with discrepant data on \n16, let alone disagreement on data
on \c12ag capture reaction, and disagreement 
in the extracted S-factor, do
not allow us to conclude that the p-wave S-factor for the \c12ag
reaction is known with an accuracy sufficient for stellar 
evolution models, and we do not confirm neither the TRIUMF data nor
the large S-factor quoted by TRIUMF with a relatively high accuracy.

\section{Acknowledgments}

I would like to acknowledge the work of my graduate student at
Yale University, Dr. Ralph H. France III now at 
UConn, on the \n16 data.

\end{document}